\documentclass[10pt,aps,prl,twocolumn,amsmath,amssymb,superscriptaddress,floatfix,footinbib,nobibnotes]{revtex4-2}

\usepackage[utf8]{inputenc}
\usepackage[english]{babel}
\usepackage{graphicx}
\usepackage{amsmath}
\usepackage{amssymb}
\usepackage{amsfonts}
\usepackage{mathtools}
\usepackage{xcolor}
\usepackage[hidelinks,colorlinks=true,linkcolor=blue,citecolor=blue,urlcolor=blue]{hyperref}

\graphicspath{{pictures/}{../}}

\newcommand{\Diss}{\mathcal{D}}

\newcommand{\gzero}{g^{(2)}(0)}
\newcommand{\sout}{s_{\rm out}}
\newcommand{\sinp}{s_0}
\newcommand{\Sopt}{S_{\rm opt}}
\newcommand{\GammaTot}{\Gamma_{\rm tot}}
\DeclareMathOperator{\sgn}{sgn}

\begin{document}

\title{Bell-inequality violation in light transmitted through disordered emitter ensembles}

\affiliation{Institute for Theoretical Physics, University of Innsbruck, Technikerstr. 21a, 6020 Innsbruck, Austria}
\affiliation{Institute for Quantum Optics and Quantum Information of the Austrian Academy of Sciences, 6020 Innsbruck, Austria}
\affiliation{Institute for Theoretical Physics, Leibniz Universit{\"a}t Hannover, Appelstra{\ss}e 2, 30167 Hannover, Germany}

\author{Ruolin Guan}
\thanks{These authors contributed equally.}
\affiliation{Institute for Theoretical Physics, Leibniz Universit{\"a}t Hannover, Appelstra{\ss}e 2, 30167 Hannover, Germany}

\author{Vineesha Srivastava}
\thanks{These authors contributed equally.}
\affiliation{Institute for Quantum Optics and Quantum Information of the Austrian Academy of Sciences, 6020 Innsbruck, Austria}
\affiliation{Institute for Theoretical Physics, University of Innsbruck, Technikerstr. 21a, 6020 Innsbruck, Austria}

\author{Kasper J. Kusmierek}
\affiliation{Institute for Theoretical Physics, Leibniz Universit{\"a}t Hannover, Appelstra{\ss}e 2, 30167 Hannover, Germany}

\author{Ivan Vybornyi}
\affiliation{Institute for Theoretical Physics, Leibniz Universit{\"a}t Hannover, Appelstra{\ss}e 2, 30167 Hannover, Germany}

\author{Klemens Hammerer}
\email{Klemens.Hammerer@uibk.ac.at}
\affiliation{Institute for Theoretical Physics, University of Innsbruck, Technikerstr. 21a, 6020 Innsbruck, Austria}
\affiliation{Institute for Quantum Optics and Quantum Information of the Austrian Academy of Sciences, 6020 Innsbruck, Austria}
\affiliation{Institute for Theoretical Physics, Leibniz Universit{\"a}t Hannover, Appelstra{\ss}e 2, 30167 Hannover, Germany}

\date{\today}

\begin{abstract}
We show that light transmitted through a disordered ensemble of weakly coupled two-level emitters can violate a Bell inequality in a continuous-wave Franson-type measurement.  The Bell signal is determined by two steady-state output-field correlations, the equal-time intensity correlation \(g^{(2)}(0)\) and the phase-sensitive two-photon coherence \(R\).  We compute these quantities for bidirectional propagation through disordered emitter ensembles using a fourth-order cumulant expansion of the many-body master equation.  The resulting Bell inequality violation appears in two distinct regimes, an antibunched regime where equal-time coincidences are suppressed and a bunched regime where photon pairs remain phase coherent.  Extrapolating the numerics to a representative weak single-emitter coupling \(\beta=0.01\) predicts violation in the antibunched regime for atom numbers \(N\simeq75\)--\(220\), and in the bunched regime for \(N\gtrsim250\).
\end{abstract}

\maketitle

Collective scattering can convert weak single-emitter nonlinearities into strong correlations of a transmitted optical mode~\cite{roy2017colloquium,sheremet2023waveguide}.  This has been analyzed and demonstrated in nanofiber-coupled atomic ensembles, where resonant transmission can produce antibunched or bunched light and where squeezing spectroscopy and two-photon interference reveal a phase-sensitive correlated two-photon component~\cite{mahmoodian2018strongly,prasad2020correlating,hinney2021unraveling,cordier2023tailoring}.  Similar physics can be expected in high-optical-depth free-space clouds, as suggested by both theory and experiments on cooperative modifications of light transmission and scattering in extended atomic media~\cite{jennewein2016coherent,zhu2016light,ferioli2023nonequilibrium,agarwal2024directional,goncalves2025phase}.

Franson interferometry provides a way to turn photon correlations into a Bell-inequality test~\cite{franson1989bell,kwiat1993high,tittel1998violation}.  Related energy-time Bell-test proposals and two-level-emitter Franson experiments have been discussed in Refs.~\cite{cabello2009proposed,peiris2017franson}.  Bell violations have recently been demonstrated with light scattered by individual two-level emitters~\cite{liu2024violation,wang2025purcell,wang2026engineering}.  These single-emitter implementations rely on strong light-matter coupling and use spectral filtering or interferometric preparation to isolate the nonlinear two-photon component.

Here we ask whether collective scattering from many weakly coupled emitters can provide a continuous-wave source for the same Franson-type Bell test using the transmitted steady-state field itself.  In the stationary-field formulation used here, the phase-optimized CHSH parameter can be written in terms of two output-field moments: the equal-time intensity correlation \(g^{(2)}(0)\) and the phase-sensitive two-photon coherence \(R=|\langle a^2\rangle|^2/\langle(a^\dagger)^2a^2\rangle\).  Determining the Bell signal therefore requires access to both the photon-pair intensity and the coherent pair amplitude of the transmitted field.  We compute these quantities for bidirectional propagation through disordered ensembles~\cite{kusmierek2025emergence} using a fourth-order cumulant expansion of the many-body master equation~\cite{kramer2015generalized,kusmierek2023higher,PhysRevA.104.023702,eltohfa2026photonstatisticschiralwaveguide}.

We find Bell-inequality violations in both antibunched and phase-coherent bunched regimes.  We also determine the transmitted photon flux and the requirements on the single-emitter coupling \(\beta\) and collective optical depth \(D=4\beta N\).  Extrapolating to the representative weak single-emitter coupling \(\beta=0.01\), where direct fourth-order calculations are prohibitive, gives threshold atom numbers \(N\simeq75\)--\(220\) for the antibunched regime and \(N\gtrsim250\) in the bunched regime.

\begin{figure}[!t]
    \centering
    \includegraphics[width=\columnwidth]{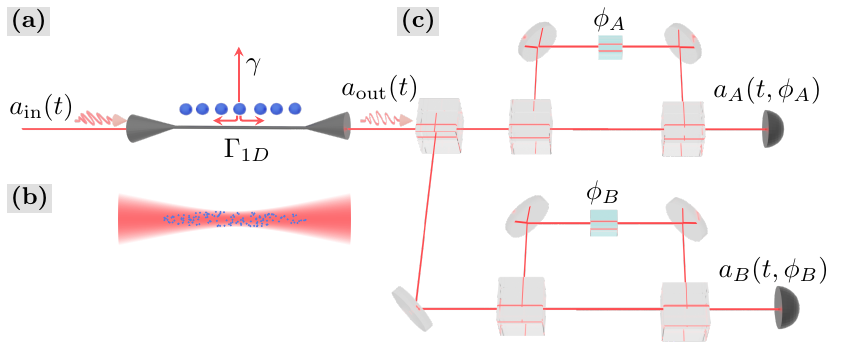}
    \caption{A coherent continuous-wave field \(a_{\mathrm{in}}(t)\) drives, from the left, a disordered ensemble of two-level systems coupled either (a) to a waveguide or (b) in free space. The atoms couple weakly to the left- and right-propagating 1D fields and strongly to environmental modes, with corresponding scattering rates \(\Gamma_{1D}\ll\gamma\). (c) The field transmitted to the right, \(a_{\mathrm{out}}(t)\), is split and analyzed with two matched unbalanced Mach-Zehnder interferometers. The phases \(\phi_A\) and \(\phi_B\) define the local settings of the Bell test on the detected fields \(a_{\mu}(t,\phi_\mu)\), with \(\mu=A,B\).}
    \label{fig:franson_setup}
\end{figure}


\label{sec:franson}

We now define the two-interferometer Franson setup for a stationary field, following Liu \emph{et al.}~\cite{liu2024violation}.  As shown in Fig.~\ref{fig:franson_setup}(c), a stationary transmitted field \(a(t)\equiv a_{\rm out}(t)\) is split and sent to two unbalanced Mach-Zehnder interferometers, labeled \(A\) and \(B\), each with delay \(T=L/c\) and adjustable phase \(\phi_\mu\).  We write the field at the \(+\) port of interferometer \(\mu=A,B\) as \(a_\mu(t,\phi_\mu)=[a(t)+e^{-i\phi_\mu}a(t-T)]/\sqrt{2}\).  The \(-\) port at phase \(\phi_\mu\) gives the same signal as the \(+\) port at phase \(\phi_\mu+\pi\), so we restrict to the \(+\) port without loss of generality.  For an arbitrary relative detection time \(\tau\), the two-interferometer coincidence function is \(G^{(2)}(\phi_A,\phi_B;\tau)=\langle a_A^\dagger(t+\tau,\phi_A)a_B^\dagger(t,\phi_B)a_B(t,\phi_B)a_A(t+\tau,\phi_A)\rangle\).  Because each interferometer has a short and a long arm, the coincidence histogram has a central peak from the indistinguishable short-short and long-long alternatives at \(\tau=0\), and two side peaks from the distinguishable short-long and long-short alternatives at \(\tau=\pm T\).  In the following we restrict to the central-peak coincidences around \(\tau=0\).  We assume that \(T\) exceeds the first-order coherence time of the field, so unequal-time coherences are negligible.

\begin{figure*}[!t]
    \centering
    \includegraphics[width=0.9\textwidth]{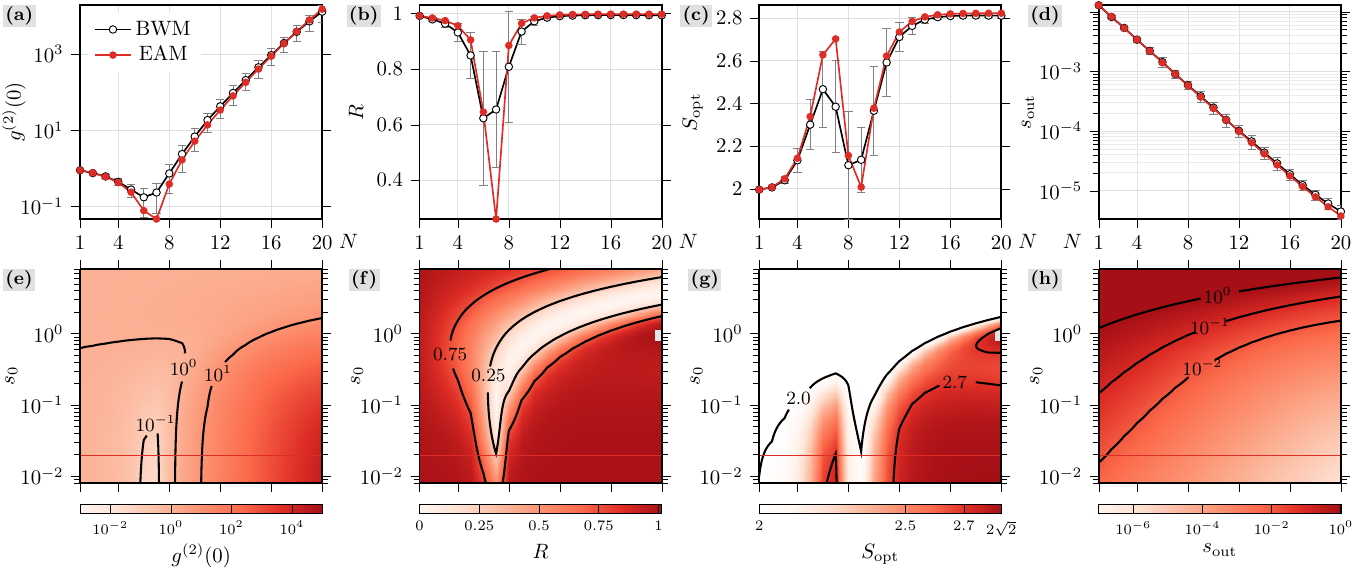}
    \caption{(Top panels) Disorder-averaged steady-state observables for fixed-position realizations of the bidirectional waveguide model (BWM) and comparison with the ensemble-averaged model (EAM).  The panels \((a) - (d)\) respectively show \(\gzero\), \(R\), \(\Sopt\), and  \(\sout\) versus emitter number \(N\) for coupling strength \(\beta=0.1\), disorder \(\eta=0.5\), and fixed input saturation \(\sinp=0.0198\).  Black circles show the average over 20 randomly chosen configurations with standard-deviation error bars, and red circles show the EAM result. (Bottom panels) Ensemble-averaged model at disorder \(\eta=0.5\). The panels \((e) - (h)\) respectively show the fourth-order cumulant maps of \(\gzero\), \(\sout\), \(R\), and \(\Sopt\) versus \(N\) and \(\sinp\) for \(\beta=0.1\). Red lines correspond to the cuts shown in \((a) - (d)\). Grey boxes in 
    \((f,g)\) denote data points where fourth order cumulant expansion yields nonphysical values of \(R\) exceeding unity by about 1\%.}
    \label{fig:realizations_eam}
\end{figure*}

For fixed phases we define the four central-peak coincidence rates by \(
    C_{mn}(\phi_A,\phi_B)
    =
    G^{(2)}(\phi_A+m\pi,\phi_B+n\pi;0)\),
where \(m,n\in\{0,1\}\). Conditioned on a central-peak coincidence, these rates define probabilities \(p_{mn}=C_{mn}/\sum_{m',n'=0}^1 C_{m'n'}\).  For each choice of interferometer phases, we assign the binary outcome values \((-1)^m\) and \((-1)^n\) to the two output ports of interferometers \(A\) and \(B\), respectively.  The corresponding correlator is
\begin{align}
    E(\phi_A,\phi_B)
    &=
    \sum_{m,n=0}^1 (-1)^{m+n}p_{mn}(\phi_A,\phi_B)
    \label{eq:E_from_C}
\end{align}
For two phase settings on each side, \((\phi_A,\phi_A')\) and \((\phi_B,\phi_B')\), the standard CHSH construction~\cite{clauser1969proposed} gives
\begin{align}
    S
    =
    \left|
    E(\phi_A,\phi_B)
    +E(\phi_A,\phi_B')
    -E(\phi_A',\phi_B)
    +E(\phi_A',\phi_B')
    \right|.
    \label{eq:CHSH}
\end{align}
Any local-realistic theory obeys the Bell inequality (BI) \(S\le2\), while quantum mechanics allows \(S\le2\sqrt{2}\), which is the Tsirelson bound.
The construction of \(S\) follows Ref.~\cite{liu2024violation}; we next express it in terms of steady-state correlations and in a second step apply it to the transmitted field from a disordered emitter ensemble.

For fixed output-field statistics, the interferometer phases only choose the local analyzer bases.  Optimizing Eq.~\eqref{eq:CHSH} over these phases gives the phase-optimized CHSH value
\begin{align}
    \Sopt
    =
    2\sqrt{2}\,
    \frac{\sqrt{R^2\gzero^2+1}}
         {1+\gzero},
    \label{eq:Sopt}
\end{align}
where the equal-time normalized intensity correlation is \(\gzero=\langle (a^\dagger)^2a^2\rangle/\langle a^\dagger a\rangle^2\).  The phase-sensitive two-photon coherence is \(R=|\langle a^2\rangle|^2/\langle (a^\dagger)^2a^2\rangle\), which compares the coherent pair amplitude with the total equal-time pair intensity.  By Cauchy-Schwarz, \(0\le R\le1\).  A strongly bunched field can therefore still have \(R\simeq1\), in which case the photon pairs have a well-defined two-photon phase; thermal or incoherent bunching instead has small \(R\).  Finally, the transmitted photon flux is \(P_{\rm out}=\langle a^\dagger a\rangle\).  With these definitions the denominator of \(R\) is \(\langle(a^\dagger)^2a^2\rangle=\gzero P_{\rm out}^2\).

Eq.~\eqref{eq:Sopt} shows that BI violation is determined jointly by the equal-time intensity correlation \(\gzero\) and the phase-sensitive two-photon coherence \(R\).  For example, for a state with \(\gzero=1\), one obtains \(\Sopt=\sqrt{2}\sqrt{R^2+1}\le2\), which precludes a BI violation.  For antibunched light, \(0<\gzero\ll1\), Eq.~\eqref{eq:Sopt} gives \(\Sopt\simeq2\sqrt{2}[1-\gzero+\mathcal{O}(\gzero^2)]\), where the leading correction is independent of \(R\).  Strong antibunching can therefore drive the optimized CHSH parameter close to the Tsirelson bound, provided the transmitted flux is not too small for a practical measurement.  For bunched light, \(\gzero\gg1\), one obtains \(\Sopt\simeq2\sqrt{2}R[1-1/\gzero+\mathcal{O}(\gzero^{-2})]\).  Thus, also in a bunched regime, the BI can be violated provided the field has a large degree of phase-sensitive two-photon coherence.  It is straightforward to verify that squeezed vacuum realizes this limiting case: in the weak-squeezing limit, \(R\to1\), \(\gzero\to\infty\), and \(\Sopt\) approaches the Tsirelson bound.


We will show that BI violation occurs in both regimes for light transmitted through a disordered ensemble of weakly coupled two-level emitters.  Its correlations are obtained from the emitter dynamics and input-output relations.  The microscopic source model is the bidirectional waveguide model (BWM), in which the emitters couple to the left- and right-propagating fields of a 1D channel and are driven resonantly from the left.  The emitter density matrix obeys
\begin{align}
    \dot\rho
    &=
    -i[H,\rho]
    +\sum_{i,j=1}^N\frac{\Gamma_{ij}}{2}
    \left(2\sigma_j^-\rho\sigma_i^+
    -\{\sigma_i^+\sigma_j^-,\rho\}\right),
    \label{eq:generic_meq}
\end{align}
with \(H=\frac{1}{2}\sum_i(\Omega_i\sigma_i^-+\Omega_i^*\sigma_i^+)+\frac{1}{2}\sum_{i,j}G_{ij}\sigma_i^+\sigma_j^-\), \(\Gamma_{ij}=\gamma\delta_{ij}+\Gamma_{1{\rm D}}\cos(k_0|z_i-z_j|)\), \(G_{ij}=\Gamma_{1{\rm D}}\sin(k_0|z_i-z_j|)\), and \(\Omega_i=2\sqrt{\Gamma_{1{\rm D}}/2}\sqrt{P_{\rm in}}\,e^{-ik_0z_i}\).  Here \(z_i\) is the position of emitter \(i\), \(\Gamma_{1{\rm D}}\) is the total decay rate into the selected 1D channel, \(\gamma\) is the spontaneous-emission rate into all other modes, and \(P_{\rm in}\) is the input photon flux.  As sketched in Fig.~\ref{fig:franson_setup}(a), the BWM is a microscopic model for emitters coupled to a bidirectional guided mode.  For the free-space or paraxial geometry of Fig.~\ref{fig:franson_setup}(b), we use the same BWM as an effective 1D description of the selected transmitted mode.

Disorder is introduced through \(z_{j+1}-z_j=n\lambda/2+\delta z_j\), where \(n\in\mathbb{N}\) and the \(\delta z_j\) are independent Gaussian variables with standard deviation \(\eta\lambda/2\).  We either solve Eq.~\eqref{eq:generic_meq} for fixed position realizations and average the resulting observables, or average the master equation itself to obtain an ensemble-averaged model (EAM)~\cite{kusmierek2025emergence}.  The average can be performed in the chiral gauge, \(\sigma_i^\pm e^{\pm ik_0z_i}\mapsto\sigma_i^\pm\), in which the master equation depends on relative positions and, after averaging, only on \(q_{ij}=\langle e^{-2ik_0(z_i-z_j)}\rangle=e^{-2(\eta\pi)^2|i-j|}\). In chiral gauge and ensemble average, Eq.~\eqref{eq:generic_meq} keeps its structure with \(H=\sqrt{\sinp/8}\sum_i(\sigma_i^-+\sigma_i^+)+\frac{1}{2}\sum_{i,j}G_{ij}\sigma_i^+\sigma_j^-\), \(\Gamma_{ij}=\delta_{ij}+\beta(1-\delta_{ij})(1+q_{ij})\), and \(G_{ij}=i\beta(1-\delta_{ij})[q_{ij}\sgn(i-j)+\sgn(j-i)]\). We use here a rescaled  time \(\GammaTot t\), with the total decay rate \(\GammaTot=\gamma+\Gamma_{1{\rm D}}\).  We also introduce the single-emitter coupling \(\beta=\Gamma_{1{\rm D}}/(2\GammaTot)\), and the input and output saturation parameters \(\sinp=8P_{\rm in}/P_{\rm sat}\) and \(\sout=8P_{\rm out}/P_{\rm sat}\), which express the corresponding photon fluxes in units of \(P_{\rm sat}=\GammaTot/\beta\), the saturation photon flux of the two-level transition~\footnote{Saturation corresponds to \(s_{\rm in}=2\Omega^2/\Gamma_{\rm tot}^2\simeq 1\); using \(\Omega=\sqrt{2\Gamma_{1D}P_{\rm in}}\) gives \(P_{\rm sat}=\Gamma_{\rm tot}/\beta\).}. In the limit of large disorder, $\eta\gg 1$, the EAM reduces to the unidirectional or chiral model~\cite{kusmierek2025emergence}.  The explicit EAM, single-realization, and chiral-limit equations are collected in the Supplementary Material below.

In the chiral gauge, the transmitted field to the right of the ensemble is \(a\equiv a_{\rm out}=a_{\rm in}-i\sqrt{\Gamma_{1{\rm D}}/2}\,S^-\), with \(S^-=\sum_j\sigma_j^-\) and \(\langle a_{\rm in}\rangle=\sqrt{P_{\rm in}}\).  We evaluate all moments entering \(\Sopt\), \(\gzero\), \(R\), and \(P_{\rm out}\) from the steady-state spin correlations.  Since \(\gzero\) contains the fourth-order moment \(\langle(S^+)^2(S^-)^2\rangle\), we use a fourth-order cumulant expansion, setting higher-order cumulants to zero~\cite{kramer2015generalized,kusmierek2023higher,kusmierek2025emergence}.  This gives \(\sum_{m=1}^4 3^m\binom{N}{m}\) coupled equations, which are integrated to steady state.  The input-output moment formulas are given in the Supplementary Material. The results were benchmarked against exact QuTiP calculations at small \(N\)~\cite{johansson2012qutip}.


We first compare the EAM to the average over single-position realizations of BWM.  Figure~\ref{fig:realizations_eam} (a)-(d) shows steady-state results for \(\gzero\), \(R\), \(\Sopt\), and \(\sout\), averaged over twenty different realizations.  For the fixed input photon flux \(\sinp=0.0198\), 1D coupling strength \(\beta=0.1\), and disorder \(\eta=0.5\), BI violations occur in both the antibunched and bunched regimes.  The comparison shows that the EAM captures the disorder-averaged behavior of the relevant Bell observables, while realization-to-realization fluctuations are largest in the nonlinear crossover region where \(R\) is suppressed and \(\Sopt\) changes rapidly.  We therefore use the EAM as it allows to more efficiently achieve extended parameter scans.

\begin{figure*}[!t]
    \centering
    \includegraphics[width=\textwidth]{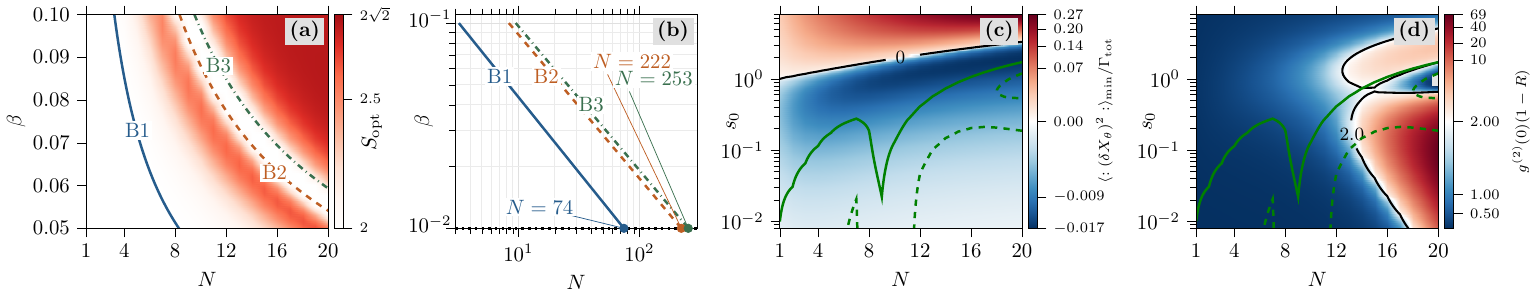}
    \caption{(a,b) Optimal Bell parameter \(\Sopt\) versus \(N\) and \(\beta\) for input saturation \(s_0 = 0.08\) and disorder \(\eta = 0.5\), and extrapolation of the \(\Sopt=2\) contour branches toward smaller \(\beta\). The contour branch fits in (b) are of the form \(\beta = c N^{-\alpha}\), with \( (c, \alpha) = (0.2358, 0.7334)\),  \( (0.4429, 0.7017)\), and \( (0.4864, 0.7017) \) for B1, B2, and B3, respectively. The exponent for B2 and B3 was enforced to be the same. (c) Minimum equal-time normally ordered quadrature variance \(\min_\theta\langle:(\delta X_\theta)^2:\rangle/\Gamma_\mathrm{tot}\) versus \(N\) and \(s_0\) for coupling strength \(\beta =0.1\) and disorder \(\eta=0.5\). (d) Map of \(g^{(2)}(0)(1-R)\) in the same parameter plane as (c). The black contour marks the Gaussian bound \(g^{(2)}(0)(1-R)=2\); values above the black contour rule out a Gaussian description of the equal-time correlations. Plots (c) and (d) are plotted in the same parameter plane as Fig.~\ref{fig:realizations_eam}(g), and green solid and dashed contours indicate \(\Sopt=2\) and \(\Sopt=2.7\), respectively. The grey box in \((d)\) denotes data points where fourth order cumulant expansion yields nonphysical values of \(R\) exceeding unity by about 1\%, cf. Fig.~\ref{fig:realizations_eam}(f).}
    \label{fig:eam_combined}
\end{figure*}

We now discuss the ensemble-averaged model at disorder \(\eta=0.5\) and strength of 1D coupling \(\beta=0.1\).  Figure~\ref{fig:realizations_eam} (e)-(h) collects steady-state maps of \(\gzero\), \(R\), \(\Sopt\), and \(\sout\) as functions of emitter number \(N\) and input saturation \(\sinp\).  Together these maps contain two distinct violation regions separated by a narrow low-\(R\) trench. The first violation mechanism is an antibunched window at weak drive and intermediate emitter number.  In this region \(\gzero<1\), and in the deepest part of the valley \(\gzero\) can be much smaller than unity with \(\Sopt\) approaching Tsirelson's bound. The island where the CHSH inequality is violated is rather narrow in \(N\) and \(\sinp\). The second mechanism appears at larger \(N\).  There the output becomes strongly bunched, with \(\gzero\gg1\), but \(R\) recovers to values close to unity.  According to the large-\(\gzero\) expansion above, this is precisely the condition for a bunched field to violate the BI.   The disordered ensemble produces phase-coherent bunching: many equal-time photon pairs are emitted with a stable two-photon phase, so the Franson interferometer can convert the pair coherence into central-peak interference.

The \(s_{\rm out}\) flux map in Fig.~\ref{fig:realizations_eam} shows the cost of operating in these regimes.  The strongest \(\Sopt\) values occur at low \(\sinp\), where \(\sout\) is small.  Increasing the drive improves brightness but tends to wash out the violation, either by increasing saturation noise or by moving the system through the low-\(R\) trench.  A useful operating point therefore lies near the boundary of the broad large-\(N\) violation region, where \(R\) has recovered, \(\Sopt>2\), and \(\sout\) has not yet become prohibitively small.

The optimized CHSH map in Fig.~\ref{fig:eam_combined}(a) shows the same violation structure in a different parameter plane: versus \(N\) and \(\beta\) at fixed \(\sinp=0.08\).  The boundary of BI violation is controlled mainly by the accumulated collective nonlinearity, which grows with both coupling and emitter number.  For larger \(\beta\), fewer emitters are needed to enter the phase-coherent bunched regime.  For weaker coupling, the violation shifts to larger \(N\).  The diagonal non-violation trench follows the suppression of \(R\), while the high-\(N\), high-\(\beta\) side of the map reaches \(\Sopt\) close to \(2\sqrt{2}\). The scaling of the violation boundary with coupling can be estimated from the \(N\)-dependence of the contour branches.  Figure~\ref{fig:eam_combined} (b) shows power-law fits to the \(\Sopt=2\) branches and extrapolates them to smaller \(\beta\), where numerics becomes prohibitive due to the large number of particles, using \(\beta=0.01\) as a reference value for state of the art experiments~\cite{prasad2020correlating}.  This extrapolation is only a guide to the emitter numbers required at weaker coupling; it assumes that the fitted branch scaling remains valid outside the calculated parameter window.  The fits indicate that the narrow low-\(N\) violation would start around \(N\simeq75\), while access to the broader bunched regime requires a few hundred emitters, roughly \(N\simeq2.5\times10^2\). These two regimes give complementary experimental strategies.  The antibunched window uses the scarcity of equal-time coincidences as the resource, but it is narrow and dim.  The bunched regime uses the collective emission of photon pairs, and can be broader in parameter space, but it requires that the pair amplitude remain coherent.

The bunched regime raises the question whether the violation is simply the response of Gaussian squeezed light.  Figure~\ref{fig:eam_combined}(c) plots the normally ordered quadrature variance, minimized with respect to the quadrature angle \(\theta\), \(\min_\theta\langle:(\delta X_\theta)^2:\rangle/\Gamma_{\rm tot}\), with \(X_\theta=(ae^{-i\theta}+a^\dagger e^{i\theta})/2\) and \(\delta X_\theta=X_\theta-\langle X_\theta\rangle\).  Negative values indicate squeezing below the vacuum level, and the plot therefore shows that the bunched Bell-violating region is weakly quadrature squeezed.  However, quadrature squeezing alone does not imply Gaussian statistics.  For a Gaussian state of the transmitted field, Wick factorization fixes the normally ordered quartic moment \(\langle(a^\dagger)^2a^2\rangle\) in terms of first- and second-order moments. Together with \(P_{\rm out}=\langle a^\dagger a\rangle\) and \(R=|\langle a^2\rangle|^2/\langle(a^\dagger)^2a^2\rangle\), this implies \(\gzero(1-R)=[\langle(a^\dagger)^2a^2\rangle-|\langle a^2\rangle|^2]/P_{\rm out}^2\le2\), which is a necessary condition for Gaussian statistics, as shown in the Supplementary Material.  Values above 2 therefore rule out a Gaussian description of the equal-time correlations. A similar Wick-factorization test has recently been used for connected third-order correlations~\cite{wang2025nongaussian}. Figure~\ref{fig:eam_combined}(d) shows that, within the bunched BI-violating region of Fig.~\ref{fig:realizations_eam}(g), the lobe connected to smaller input drive lies in the domain \(\gzero(1-R)>2\).  For this lobe, the BI violation cannot be explained by Gaussian squeezing of the transmitted mode alone; it requires non-Gaussian pair correlations generated by the emitter nonlinearity. For the lobe near \(s_0\simeq1\), however, \(\gzero(1-R)\) stays below the Gaussian bound, so this criterion does not rule out a Gaussian description.  At the same time, the cumulant solution begins to produce slightly nonphysical values with \(R>1\), by about \(1\%\), indicating sensitivity to correlations beyond the fourth-order truncation.  This suggests that higher-order field correlations may play an important role in this regime, independently of the Gaussianity question.


In conclusion, we have shown that light transmitted through a disordered ensemble of weakly coupled two-level emitters can violate a Bell inequality in a continuous-wave Franson-type test, with the optimized CHSH parameter determined by \(\gzero\) and \(R\).  The violation occurs in two regimes: an antibunched window where suppressed equal-time coincidences drive \(\Sopt\) upward, and a bunched regime where phase-coherent, non-Gaussian pair correlations keep \(R\) large.  An important next step is to go beyond the equal-time, infinitesimal-gate approximation and include the finite acceptance window of the central coincidence peak; this requires two-time, and generally time-integrated, output-field correlations, which are substantially more challenging within the present cumulant approach.  More broadly, arrays of Franson interferometers and multipartite time-bin Bell inequalities may reveal an intriguing connection between higher-order field correlation functions and BI violations, extending the role played here by \(\gzero\) and \(R\).

\begin{acknowledgments}
We acknowledge discussions with Anders S.~S{\o}rensen, Philipp Schneeweiss, and Arno Rauschenbeutel.  We acknowledge support from DFG through the Collaborative Research Center SFB1227 (DQ-mat, Project-ID 274200144), from the Federal Ministry for Research, Technology and Space (BMFTR), Germany, through project ATIQ, and from Quantum Valley Lower Saxony (QVLS), funded by the Volkswagen Foundation and the Ministry for Science and Culture of Lower Saxony.
\end{acknowledgments}

\bibliography{bib}

\newpage
\pagebreak

\onecolumngrid

\vfill
\pagebreak

\setcounter{equation}{0}
\setcounter{figure}{0}
\setcounter{table}{0}
\setcounter{section}{0}
\renewcommand{\theequation}{S\arabic{equation}}
\renewcommand{\thefigure}{S\arabic{figure}}

\center{\large{\textbf{Supplementary Material}}}

\raggedright

\section{Dimensionless chiral-gauge models}
\label{app:models}

In this section, we present in detail the dimensionless quantum master equations derived in the chiral gauge for the ensemble average model (EAM) and the bidirectional waveguide model (BWM) for single positional realizations, used in the main text.
The main text writes the fixed-position dynamics first in the normal gauge, Eq.~\eqref{eq:generic_meq} of the main text, and gives the EAM coefficients in the dimensionless chiral gauge.  Formally, the laser phase is absorbed into the atomic operators,
\begin{align*}
    \sigma_i^\pm e^{\pm ik_0z_i}\mapsto\sigma_i^\pm .
\end{align*}
In this convention the EAM master equation is
\begin{align}
    \frac{d\rho}{d(\GammaTot t)}
    &=
    -i[H,\rho]
    +\sum_{i,j=1}^N\frac{\Gamma_{ij}}{2}
    \left(2\sigma_j^-\rho\sigma_i^+
    -\{\sigma_i^+\sigma_j^-,\rho\}\right),
    \label{eq:eam_meq_appendix}
\end{align}
where, with \(q_{ij}=e^{-2(\eta\pi)^2|i-j|}\),
\begin{align}
    H
    &=
    \sqrt{\frac{\sinp}{8}}
    \sum_{i=1}^N(\sigma_i^-+\sigma_i^+)
    +\frac{1}{2}\sum_{i,j}G_{ij}\sigma_i^+\sigma_j^-,
    \nonumber\\
    \Gamma_{ij}
    &=
    \delta_{ij}
    +\beta(1-\delta_{ij})(1+q_{ij}),
    \nonumber\\
    G_{ij}
    &=
    i\beta(1-\delta_{ij})
    \left[q_{ij}\sgn(i-j)+\sgn(j-i)\right].
    \label{eq:eam_coefficients_appendix}
\end{align}
The cumulant equations use this dimensionless chiral-gauge convention for all models.

For single-position realizations and the chiral limit, it is convenient to split off the local decay term.  After measuring time in units of \(\GammaTot^{-1}\), these bidirectional waveguide models (BWM) have the form
\begin{align}
    \frac{d\rho}{d(\GammaTot t)}
    &=
    -i[H_X,\rho]
    +\sum_{i=1}^N\Diss[\sigma_i^-]\rho
    \nonumber\\
    &\quad
    +\sum_{i\ne j}\frac{\Gamma_{ij}^{(X)}}{2}
    \left(2\sigma_j^-\rho\sigma_i^+
    -\{\sigma_i^+\sigma_j^-,\rho\}\right),
    \label{eq:chiral_meq_appendix}
\end{align}
where
\begin{align}
    H_X
    =
    \sqrt{\frac{\sinp}{8}}
    \sum_{i=1}^N(\sigma_i^-+\sigma_i^+)
    +\frac{1}{2}\sum_{i\ne j}G_{ij}^{(X)}
    \sigma_i^+\sigma_j^-,
\end{align}
Here \(\Diss[o]\rho=o\rho o^\dagger-\{o^\dagger o,\rho\}/2\), and \(X\) labels the coefficient set.  The local dissipator has unit rate in this convention.

For one positional realization \(\{z_i\}\), the chiral-gauge coefficients are
\begin{align}
    \Gamma_{ij}^{({\rm sr})}
    &=
    \beta\left[1+e^{-2ik_0(z_i-z_j)}\right],
    \qquad i\ne j,
    \label{eq:Gamma_single}\\
    G_{ij}^{({\rm sr})}
    &=
    2\beta\sin\!\left(k_0|z_i-z_j|\right)
    e^{-ik_0(z_i-z_j)},
    \qquad i\ne j.
    \label{eq:G_single}
\end{align}
These coefficients describe both propagation directions for a fixed microscopic disorder realization while keeping the coherent drive spatially uniform.  Replacing \(e^{-2ik_0(z_i-z_j)}\) by its Gaussian disorder average gives the EAM coefficients \(\Gamma_{ij}\) and \(G_{ij}\) above.

For large disorder, \(e^{-2(\eta\pi)^2|i-j|}\to0\), so the EAM reduces to the chiral or unidirectional model,
\begin{align}
    \Gamma_{ij}^{({\rm ch})}
    &=
    \beta,
    \qquad i\ne j,\\
    G_{ij}^{({\rm ch})}
    &=
    i\beta\,\sgn(j-i),
    \qquad i\ne j.
\end{align}

\section{Output moments from the collective dipole}
\label{app:moments}

In this section, we present the expressions for the normalized output flux, $s_{\rm out}$, the  equal-time second order coherence, $g^{(2)}(0)$, the phase sensitive two photon coherence, $R$, and the quadrature variance, $\langle : \delta X^2_{\theta} : \rangle$,  corresponding to the output field in terms of the collective spin moments. 

Let
\begin{align}
    S^-=\sum_{i=1}^N\sigma_i^-,
    \qquad
    a=a_{\rm in}-i\sqrt{\frac{\Gamma_{1{\rm D}}}{2}}\,S^-,
    \qquad
    \langle a_{\rm in}\rangle=\sqrt{P_{\rm in}}.
\end{align}
In the convention \(\beta=\Gamma_{1{\rm D}}/(2\GammaTot)\), \(\sinp=8P_{\rm in}/P_{\rm sat}\), and \(\sout=8P_{\rm out}/P_{\rm sat}\), the normalized output flux is
\begin{align}
    \sout
    =
    \sinp
    +2i\sqrt{2\sinp}\,\beta
    \left(\langle S^+\rangle-\langle S^-\rangle\right)
    +8\beta^2\langle S^+S^-\rangle .
    \label{eq:sout_appendix}
\end{align}
The equal-time second-order coherence is
\begin{align}
    \gzero
    &=
    \frac{\sinp^2}{\sout^2}
    \Bigg[
    1
    +\frac{\sqrt{32}\beta}{\sqrt{\sinp}}
    \left(-i\langle S^-\rangle+i\langle S^+\rangle\right)
    \nonumber\\
    &\quad
    +\frac{8\beta^2}{\sinp}
    \left(
    4\langle S^+S^-\rangle
    -\langle (S^-)^2\rangle
    -\langle(S^+)^2\rangle
    \right)
    \nonumber\\
    &\quad
    +\frac{32\sqrt{2}\beta^3}{\sinp^{3/2}}
    \left(
    -i\langle S^+(S^-)^2\rangle
    +i\langle (S^+)^2S^-\rangle
    \right)
    \nonumber\\
    &\quad
    +\frac{64\beta^4}{\sinp^2}
    \langle (S^+)^2(S^-)^2\rangle
    \Bigg].
    \label{eq:g2_appendix}
\end{align}
Finally,
\begin{align}
    R
    =
    \frac{
    \left|
    \dfrac{\sinp}{8\beta}
    -2i\sqrt{\dfrac{\sinp}{8}}\langle S^-\rangle
    -\beta\langle (S^-)^2\rangle
    \right|^2}
    {
    \gzero
    \left[
    \dfrac{\sinp}{8\beta}
    +\sqrt{\dfrac{\sinp}{8}}
    \left(-i\langle S^-\rangle+i\langle S^+\rangle\right)
    +\beta\langle S^+S^-\rangle
    \right]^2
    }.
    \label{eq:R_appendix}
\end{align}
Equations~\eqref{eq:sout_appendix}-\eqref{eq:R_appendix} make explicit why the fourth-order cumulant solution is required: \(\gzero\) contains the collective four-body moment \(\langle (S^+)^2(S^-)^2\rangle\).

Similarly, with the quadrature along $\theta$ defined with $X_\theta=(ae^{-i\theta}+a^\dagger e^{i\theta})/2$, the normally ordered quadrature variance below vacuum fluctuations is given as 
\begin{eqnarray}
    \frac{1}{\Gamma_{\rm tot}}\langle :\delta X_\theta^2: \rangle &=& \frac{\beta}{4}\Big[
    - e^{2i\theta}\Big(\langle  (S^-)^2\rangle - \langle S^-\rangle^2\Big)
    - e^{-2i\theta}\Big(\langle ( S^{+})^2\rangle - \langle  S^{+}\rangle^2\Big) \nonumber \\
    &&+ 2\Big(\langle S^{+} S^-\rangle - \langle  S^{+}\rangle\langle S^-\rangle\Big)\Big], 
    \label{eq:quadrature_squeezing}
\end{eqnarray}

where ${\rm min}_{\theta}\langle :\delta X_\theta^2: \rangle/ \Gamma_{\rm tot} < 0 $ indicates squeezing.

\section{Gaussianity criterion}
\label{app:gaussianity}

In this section, we derive the Gaussian bound used in Fig.~\ref{fig:eam_combined}(d) of the main text.  Let \(P=\langle a^\dagger a\rangle\), \(G=\langle(a^\dagger)^2a^2\rangle=\gzero P^2\), and \(M=\langle a^2\rangle\), so that \(R=|M|^2/G\) and \(\Lambda=\gzero(1-R)=(G-|M|^2)/P^2\).  For a Gaussian field write \(a=\alpha+d\), with \(\langle d\rangle=0\), \(n=\langle d^\dagger d\rangle\), and \(m=\langle d^2\rangle\).  Wick factorization gives
\begin{align}
    G
    &=
    |\alpha|^4+4|\alpha|^2n+\alpha^{*2}m+\alpha^2m^*
    +2n^2+|m|^2,\\
    |M|^2
    &=
    |\alpha^2+m|^2
    =
    |\alpha|^4+\alpha^{*2}m+\alpha^2m^*+|m|^2.
\end{align}
Thus \(G-|M|^2=4|\alpha|^2n+2n^2=2(P^2-|\alpha|^4)\), where \(P=|\alpha|^2+n\).  Therefore every Gaussian state obeys
\begin{align}
    \Lambda
    =
    2\left[1-\left(\frac{|\alpha|^2}{P}\right)^2\right]
    \le 2.
\end{align}
The inequality \(\Lambda\le2\) is therefore a necessary condition for Gaussianity of the equal-time correlations of the transmitted field.  In particular, values \(\Lambda>2\) exclude a Gaussian description.
\end{document}